\documentclass[12pt]{iopart}
\usepackage{mdframed}
\usepackage{graphicx}
\usepackage{url}
\usepackage{array}
\begin{document}

\title[Non-specular reflection by a planar resonant metasurface]{Non-specular 
reflection by a planar resonant metasurface}
\author{Sergey L Prosvirnin$^{1, 2}$, Vyacheslav V Khardikov$^{1, 2}$,  Kateryna L Domina$^2$, Olexandr A Maslovskiy$^2$,
Ludmila A Kochetova$^3$, and Vladimir V Yachin$^{1, 3}$}

\address{$^1$ Institute of Radio Astronomy, Kharkiv 61002, Ukraine }
\address{$^2$ Karazin Kharkiv National University, Kharkiv 61022, Ukraine}
\address{$^3$ Usikov Institute of Radio Physics and Electronics, Kharkiv 61115, Ukraine}
\ead{prosvirn@rian.kharkov.ua}
%\ead{khav77@gmail.com}

\begin{abstract}
An uncommon double-ray scenario of light resonant  scattering by a periodic metasurface is proposed to provide strong non-specular reflection. The metasurface is constructed as an array of silicon nanodisks placed on thin silica-on-metal substrate. A low-lossy non-specular resonant reflection for any direction and any polarization of incident wave is revealed by a numerical simulation of light scattering. The conditions for the implementation of an autocollimation scheme of scattering and the observation of non-specular reflected ray that does not lie in the incidence plane are worked out. It is shown that the change of dielectric substrate thickness may be applied to set the width of frequency band of non-specular reflection. The light intensity related to the specular and non-specular reflected ray can be controlled by changing  the angle of incidence or by the polarization of incident wave. 
\end{abstract}
\noindent{\it Keywords\/}: Resonant metasurface, non-specular reflection, autocollimation, Littrow scenario, diffraction efficiency

\pacs{42.25.Fx, 42.25.Ja, 78.67.-n}
%\submitto{\JOA}
%\maketitle

\section{Introduction}

A flat metallic mirror is a simple outstanding optical device  being used since ancient times \cite{ancient_mirrors-2006}, which is capable of changing sharply the direction of light rays or returning them back at the normal incidence over an extensive spectral bandwidth. A metallic flat mirror reflects the incident light in the specular direction with an only small decrease in  its intensity. Near the surface of the mirror, the field distribution has close to zero an electrical tangential component and a double magnetic one in a comparison with a field of incident wave.

Today, photonics requires mirrors with specific reflection selectivity depending on the light wavelength, both in amplitude and phase. The phase spatial variation along the reflection surface is provided in order to manipulate, redirect and concentrate light energy. Metasurfaces open prospects to create such specific reflectors \cite{metasurface-2016}.

Modern optics technologies give the opportunity to produce resonant non-specular and selectively reflecting metasurfaces. A reflecting metasurface is usually a planar double periodic subwavelength patterned metal-dielectric or all-dielectric layer placed on a metal substrate \cite{meta_mirrors-2014}. The thickness of a metasurface is usually very small in comparison with an electromagnetic  wavelength in free space. However, the periodic structured surface provides conditions for the excitation of different types of resonances. A spectacular metasurface response manifests in resonance reflection and absorption \cite{meng-2017}, in electromagnetic radiation enhancement by using a laser medium \cite{yablonovitch-2015}, in exotic electromagnetic field boundary values, which may be the same as on a surface of an artificial magnetic wall \cite{4666749}, and in the confinement of the intensive electromagnetic field inside the structure \cite{all-dielectric-2016}.

The excellent wavelength-selective properties of specifically functionalized metamaterial mirrors are necessary to design precise sensors \cite{absorber_for_sensor_2015, multispectral_sensing_2015, sydorchuk-2017}, in particular, biological material sensors \cite{sensors-2017}.  Another aspect of the application of resonant mirrors involves the placement of a certain quantum system which can have different energy states, in an intense field near the reflecting surface. In this case, the task is to obtain the maximum radiation intensity or light absorption. The practical application of the resonant arrays radiated one or only a few diffracted orders besides main one looks  attractive due to their outstanding features \cite{Collin_2014, Zhu:15}. 

The main factor defining the spectacular properties of a metasurface is the resonant interaction of electromagnetic waves with a patterned metallic or a dielectric layer. A special design of a metasurface unit cell can provide the opportunity to excite specific high-Q resonances known as bound states in the continuum \cite{khardikov-2010-tol, khardikov-2012-jop, zheludev-2013-Near-infrared}. In microwave region, the microstrip and waveguide reflection arrays that realised this kind of resonant transformation were firstly proposed and researched in \cite{prosvirnin-2009-jopa, Gribovsky_2014} respectively.

In this paper, by using a full-wave numerical simulation, we demonstrate for the first time the features of the uncommon resonant regime of the metasurface involving the excitation of one additional diffraction order besides the main partial wave to reach a full non-specular light reflection in an advance choosing direction. The  proposed silicon-on-metal metasurface manifests a non-specular reflection of light with a low dissipative loss. To provide the non-specular reflection for any direction of impingement and any polarization of incident wave, we propose an array of silicon disks with the excitation of a Mie-resonance that has a specific symmetry of field distribution. The levels of the light intensities that correspond to the specular and non-specular ray of the reflected field can be simply controlled by changing the angle of incidence of the initial wave. Other mean to tune intensity distribution between rays consists in variation of polarization of incident wave. 

\section{Problem statement}

\begin{figure}[!h]
\centering
\includegraphics[width=3.0in]{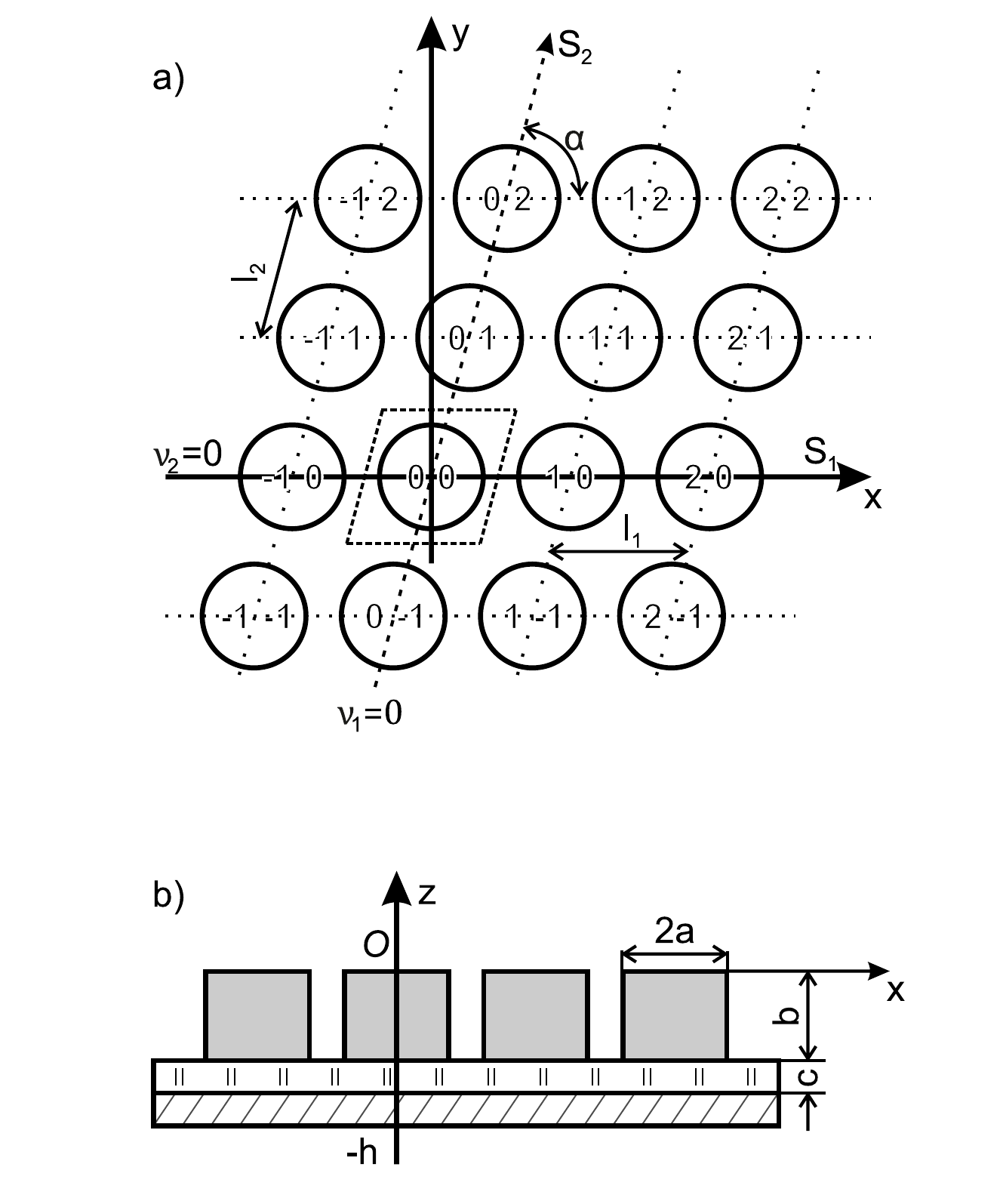}
\caption{A double-periodic planar reflect array: view in the plane $z=0$ (a), cross section in the plane $y = 0$ (b), the unit cell of array is shown by dashed lines.} \label{array}
\end{figure}

\begin{figure}
\centering
  \includegraphics[height=6cm]{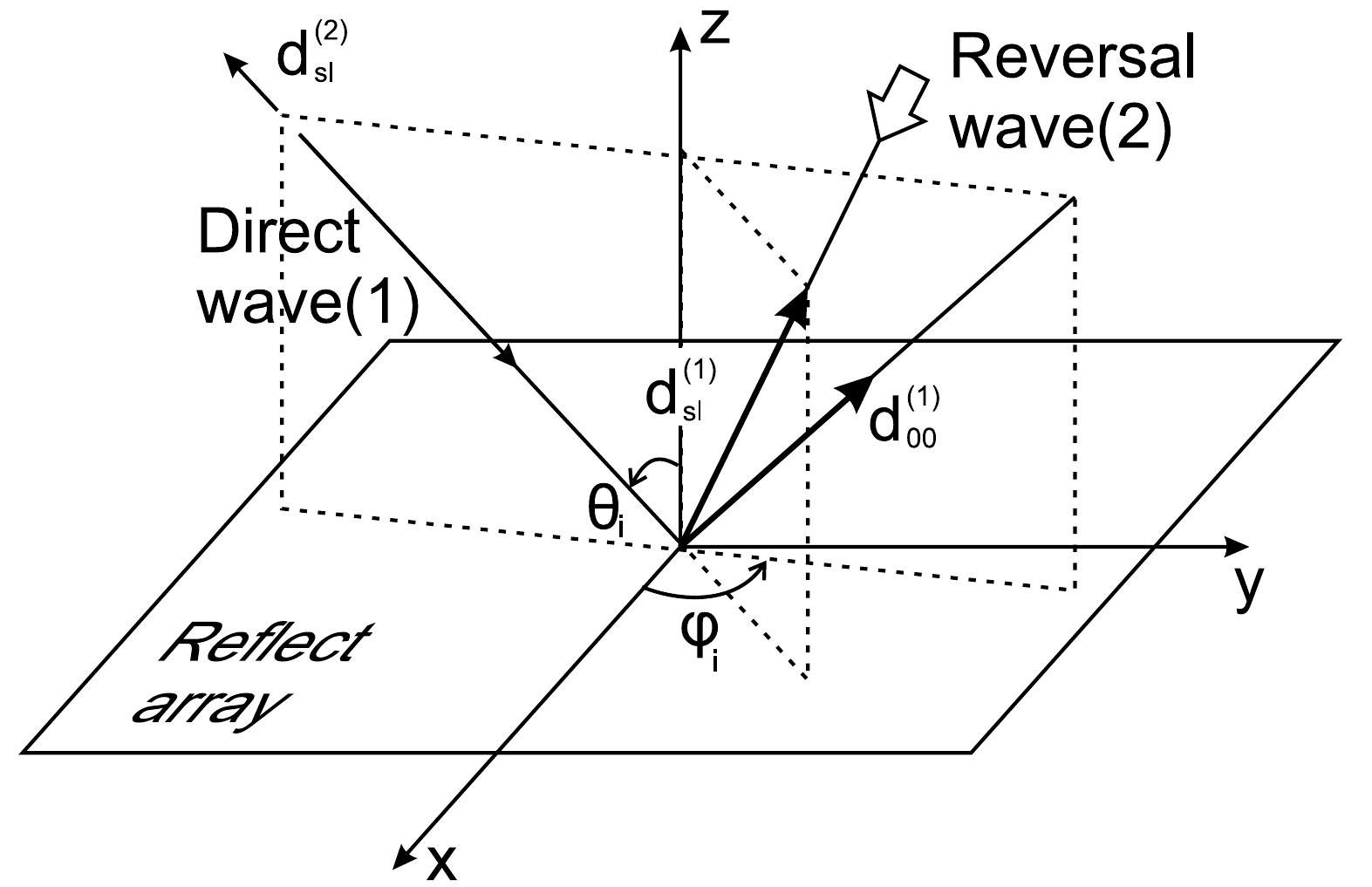}\\
  \caption{Direct and reversal scenarios of diffraction.}\label{DirRevSc}
\end{figure}

Let us consider the incidence of the plane electromagnetic wave
\begin{equation}
 \label{E1i}
  {\bi{E}}^{(1i)}={\bi e}^{(1)} \exp(-i{\bi k}^i {\bi r}) \exp(i \omega t)
\end{equation}
from region $z>0$ on a periodic structure placed in free space within the layer $-h<z<0$ parallel to the $xy$ plane (see Fig.~\ref{array}, and Fig.~\ref{DirRevSc}). Here ${\bi e}^{(1)}$ is an unit polarization vector and $\bi{k}^i$ is a wave vector. It is convenient to represent the wave vector as the sum of the components transverse (${\bi k}_\bot = {\bi e}_x k \sin\theta_i \cos\phi_i + {\bi e}_y k \sin\theta_i \sin\phi_i$) and parallel ($\gamma = (k^2-k_\bot^2)^{1/2}$) to the axis $z$, where $\phi_i$ and $\theta_i$ are the azimuth and the polar angle of the incidence direction (the angle of incidence is $\pi-\theta_i$), $k=\omega/c=2\pi/\lambda$, vectors ${\bi e}_x$, ${\bi e}_y$ and ${\bi e}_z$ are unit basis vectors along the axes $x$, $y$ and $z$. For simplicity of notation the time multiplier has been dropped below. The incident wave is assumed to be a non-evanescent plane wave.

%\textcolor{red}{To consider an infinite plane array, whose elements are periodically disposed in the mesh points plotted in the oblique coordinates $s_1$  and $s_2$  and marked by , as shown in Fig.~\ref{array}. The element position in the plane $z = 0$ is determined by a radius vector with two indices $\nu_1$ and $\nu_2$ according to}

To consider an infinite plane array, whose directions of periodicity coincide with oblique coordinates
$s_1$  and $s_2$, we number array elements by two symbols $\nu_1$ and $\nu_2$ as shown in Fig.~\ref{array}. The center of the element with $\nu_1=\nu_2=0$ lays on the axis $z$. Then the element position in the plane
$z = 0$ is determined by a radius vector with two indices $\nu_1$
and $\nu_2$ according to ${\bi{\rho}}_{\nu_1 \nu_2} = \nu_1 l_1 \bi{e}_1 + \nu_2 l_2 \bi{e}_2,$
where $\bi{e}_1$  and $\bi{e}_2$  are the unit vectors directed
towards $s_1$ and $s_2,$ and $l_1$ and $l_2$  are the
corresponding array periods. The array elements are excited by a field with constant amplitude
and linear phase shift distribution. The phase of the impinging
field at element $(\nu_1, \, \nu_2)$  is determined by the
expression $\varphi_{\nu_1 \nu_2} = (\bi{k}^i, \, \bi{\rho}_{\nu_1 \nu_2}).$

The complete set of solutions of the scalar Helmholtz equation, each
varying with the change of the coordinates $s_1$  and $s_2$ according
to Floquet's theorem, reads in the region above the periodic array
$z > 0$ \cite{Amitay-1972-tapaa}
\begin{equation}
\label{2.2}
    S_{m n} =  e^{-i ( k_1^i + 2 \pi m/l_1) s_1}
              e^{-i ( k_2^i + 2 \pi n/l_2)  s_2}
              e^{-i \gamma_{m n} z},
\end{equation}
where  $\gamma_{m n}$ is propagation constant along the $z$-axis, $k_1^i$ and $k_2^i$ are wave vector components of an incident plane wave along the $s_1$- and $s_2$-axes.

When writing expression (\ref{2.2}) in cartesian coordinates,
use the relation
$$ %\begin{eqnarray}
    \nonumber
    s_1 = x - y \cot \alpha, \quad s_2 = y / \sin \alpha
$$ %\end{eqnarray}
and represent the component of the incident wave vector along the
$s_2$-axis by its  $x$- and  $y$-components
$$ %\begin{eqnarray}
\nonumber
    k_2^i = k_x^i \cos \alpha + k_y^i \sin \alpha,
$$ %\end{eqnarray}
where  $\alpha$ is the angle between the  $O x$- and  $O
s_2$-axes. After substitution of these expressions into Eq.
(\ref{2.2}), one obtains
\begin{equation}
    \label{Smn}
    S_{m n} = e^{-i \bi{\chi}_{mn} \bi{\rho}}
              e^{-i \gamma_{m n} z},
\end{equation}
where  
\begin{equation}
\label{chi_mn}
    %\nonumber
    \bi{\chi}_{mn} = \bi{e}_x (k_x^i + \frac{2 \pi m}{l_x}) +
    \bi{e}_y ( k_y^i + \frac{2 \pi n}{l_y} - \frac{2 \pi m}{l_x \tan
    \alpha})
\end{equation}
is projection of the
propagation vector of the Floquet harmonic $S_{m n}$  on the $xOy$
plane, $\bi{\rho}=\bi{e}_x x +\bi{e}_y y$, $l_x = l_1$, $l_y = l_2 \sin \alpha$. 
Since  $S_{m n}$ is a solution of the Helmholtz equation, the
propagation constant along the $O z$-axis takes the form
$$
    \gamma_{m n} = \sqrt{k^2 - \chi_{mn}^2}, \quad \mbox{Re}
\gamma_{mn}\geq 0, \quad \mbox{Im}
\gamma_{mn}\leq 0.
$$

It is easy to verify by direct check that the space harmonics (\ref{Smn})
%$$ %\begin{eqnarray}
%    \nonumber
% S_{m n} = e^{-i ( \bi{\chi}_{mn} \bi{\rho} + \gamma_{m n} z  ) }
%$$ %\end{eqnarray}
fulfill the Floquet conditions
\begin{eqnarray}
    \nonumber
    S_{m n}(x + l_1, y, z) = S_{m n} (x, y, z)
                             e^{- i k_1^i l_1},
\end{eqnarray}
\begin{eqnarray}
    \nonumber
    S_{m n}(x + l_g, y + l_y, z) = S_{m n} (x, y, z)
                             e^{- i k_2^i l_2},
\end{eqnarray}
where $l_g = l_2 \cos \alpha.$  

Each space harmonic $S_{m n}$,
for which $\gamma_{m n}$ is real, corresponds to one of the plane
waves, which transport energy from the array plane. The harmonic
with subscripts $m = 0$  and $n = 0$ is a plane wave propagated
along a direction of incident wave specular reflection. 

The field outside of the periodic structure is a superposition of partial diffracted waves
propagating away from the array and along the array's plane:
\begin{equation}\label{E1}
  {\bi{E}}^{(1)}= 
    {\bi{E}}^{(1i)} + \sum_{m,n=-\infty}^\infty
    {\bi{d}}_{mn}^{(1)}
    \exp( - i{\bi{k}}_{mn}^{r} {\bi r}), \qquad z>0,
\end{equation}
where ${\bi{d}}_{mn}^{(1)}$ and ${\bi{k}}_{mn}^{r} =
{\bi{\chi}}_{mn} + {\bi e}_z\gamma_{mn}$ 
are amplitudes and wave
vectors of partial waves of the reflected
field. 

The diffraction of Wave~1 will be referred as
the {\it direct} diffraction scenario.

Let us choose the one of the non-evanescent waves, reflected in the direct scenario, and define for it a {\it reversed} diffraction scenario. For instance, the partial reflected wave with indexes $m=s$ and $n=l$ is considered. Then the diffraction of wave:
\begin{equation}\label{E2i1}
 {\bi{E}}^{(2\rmi)}={\bi e}^{(2)}
 \exp(\rmi{\bi{k}}_{sl}^{r} {\bi r})
\end{equation}
on studied metasurface presents the reversed scenario. The wave (\ref{E2i1}) is plane wave propagates along the direction  $-{\bi{k}}_{sl}^{r}$ from the region $z>0$ (see Fig.~\ref{DirRevSc}: Wave~2) and it is backward for reflected $\bi{d}_{sl}^{(1)} \exp( - i{\bi{k}}_{sl}^{r} {\bi r})$ wave in the direct scenario. The incident wave ${\bi{E}}^{(2\rmi)}$ creates its own diffraction pattern  ${\bi{E}}^{(2)}$, that contains the reflected partial wave propagating in the exactly opposite direction to $\bi{E}^{(1i)}$ wave. Thus, the direct and reversed diffraction scenarios definitely contains as minimum two waves which propagates in opposite directions in the same wave channel consisting of an incident wave and the $sl$ partial diffracted wave.

Let us apply the Lorentz reciprocity lemma \cite{collin-2001} to the field superposition in the direct and reversed scenarios under the assumption that the periodic structure has neither non-linear nor non-reciprocal elements. Within the volume bounded by the plane facets of the unit cell such relation may be written in the following form:
\begin{equation}\label{Lor}
  \oint_S \left\{ [{\bi{E}}^{(1)}\times {\bi{H}}^{(2)}]_{\rm N}
  -[{\bi{E}}^{(2)}\times {\bi{H}}^{(1)}]_{\rm N} \right\}{\rm d} s=0.
\end{equation}
Here $S$ is a surface enclosing the considered volume, and subindex $\rm N$ denotes the
component of the vector product, that is perpendicular to surface
$S$. By direct inspection it may be shown that the following
relation results from the Lorentz lemma \cite{prosvirnin-2009-jopa}:
\begin{equation}\label{aa1}
\gamma_{sl} \left( {\bi e}^{(2)} \cdot {\bi d}^{(1)}_{sl} \right) =
\gamma_{00}\left( {\bi e}^{(1)} \cdot {\bi d}^{(2)}_{sl} \right) 
\end{equation}
between amplitudes ${\bi d}^{(1)}_{sl}$ and ${\bi d}^{(2)}_{sl}$ of the partial waves.

An analysis of the formula (\ref{aa1}) makes it possible to reveal fundamental relation between waves that are diffracted in opposite directions in the same wave channels of a periodic structure. These relations are a consequence of the Lorentz theorem and therefore are valid for any reciprocal lossy or lossless periodic structures.
%There is particularizations of the general principle of reciprocity addressed by de Hoop \cite{potton} for the case of electromagnetic wave diffraction by planar double-periodic arrays. 
An exclusively important physical consequence of the reciprocity theorem is that a system exhibiting any extreme properties when light propagates in one direction manifests the same properties when light propagate in the opposite direction.

We have chosen dielectric disks as resonant elements (metaatoms) of the metasurface. The resonant features of single dielectric disks and finite length cylinders located in free space and on a dielectric substrate are well established  \cite{vandeGroep:13}. The attractive opportunities of using dielectric disks placed on a metal substrate as an antenna elements of the mobile devices were discussed in \cite{dielectric-antenna-2011}. The disk arrays were also studied in details in the case when their pitch is smaller than a wavelength \cite{{all-dielectric-2016}}.

The disks of designed array are placed periodically on a dielectric-on-metal plane substrate. Their axes are oriented 
orthogonally to the plane of the reflect array (see Fig. \ref{array}).  The first few modes of the free standing 
dielectric disk resonator are $\mbox{TE}_{01\delta}$, $\mbox{HEM}_{11\delta}$, 
and $\mbox{TM}_{01\delta}$ mode \cite{dielectric-antenna-2011}. If we put PEC plane orthogonally to the disk's axis and via half of its thickness, the resonant mode $\mbox{TE}_{01\delta}$ is not excited in the frequency range under interest. Two other modes mentioned above  can be excited. The hybrid mode $\mbox{HEM}_{11\delta}$ appeared most promising to be chosen as resonant mode of the metaatom of the reflect array because it can be excited from any direction by any polarized incident wave. The sketch of its field distribution is presented in the inset of Fig. \ref{map_1}. 

\section{Results of simulation and analysis}

\begin{figure}[!h]
\centering
\includegraphics[width=4.0in]{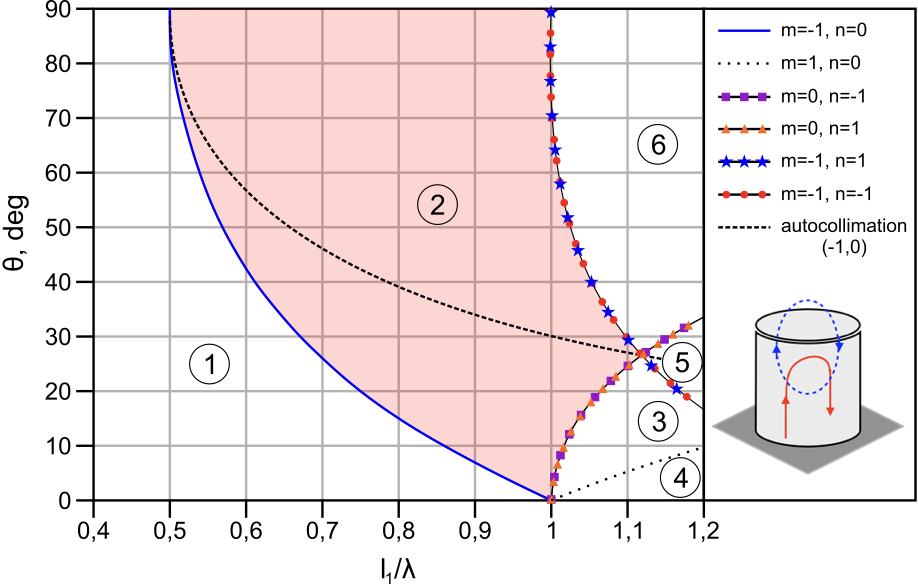}
\caption{A diagram presents dependencies of normalise cutoff frequencies of partial reflected $mn$ wave on incident polar angle and zones of diffraction problem parameters allowing the existence of different number non-evanescent diffraction orders ($l_1=l_2,$ $\alpha=\pi/2,$ $\phi_i=0.$)  The dot line crossed the zone 2 marks the Littrow diffraction scenario corresponded to (-10)-order propagation direction rigorously opposite to incident wave. A sketch of field distribution of $\mbox{HEM}_{11\delta}$ mode of dielectric resonator located on PEC substrate is shown schematically in the insert of right panel. The solid and dashed lines correspond to the electric and magnetic fields respectively.}
\label{map_1}
\end{figure}

In the assumption $\phi_i=0,$ let us consider a counter diagram presented on the plane of two parameters which are an angle of incidence $\theta_i$ and a normalize frequency $l_1/\lambda$ corresponded to different reflected orders of non-evanescent waves (see Fig. \ref{map_1}).  In the area of the diagram, marked 1, there are the parameters corresponded to the existence of only one  non-evanescent spatial diffraction harmonic of the reflected field, namely the (00)-partial wave. In the area marked 2, there is non-evanescent (-10)-wave harmonic in addition to the (00)-wave. The areas marked from 3 to 6 correspond to a larger number of partial waves going away from the reflection surface. We did not detailed this information for the sake of simplicity.

The dot line crossed the zone 2 (see Fig. \ref{map_1}) marks so-named the Littrow diffraction scenario which is important for using metasurfases as mirrors of resonators in laser applications \cite{hard-1970, Masalov-1980, Lotem-1994} and impulse compressor devices \cite{Gribovsky_2014}. In the Littrow diffraction scheme (or autocollimation diffraction regime), the (-10)-order propagation direction is rigorously opposite to direction of incident wave.

Let us consider the diffraction of TE- and TM-polarized waves by the reflect array of disk dielectric particles placed on the double layer substrate. The substrate consists of a thin silica layer located between an array of silicon disks and a metal background. For the sake of the analysis simplicity, we assume that the metal background is a perfect electric conductor (PEC). Relative permittivity of silicon is assumed $\epsilon_{Si}=11.9,$ and 
$\tan \delta_{Si} =0.001.$ Silica permittivity is approximated below as 2.1.

\begin{figure}
\centering
  \includegraphics[height=4in]{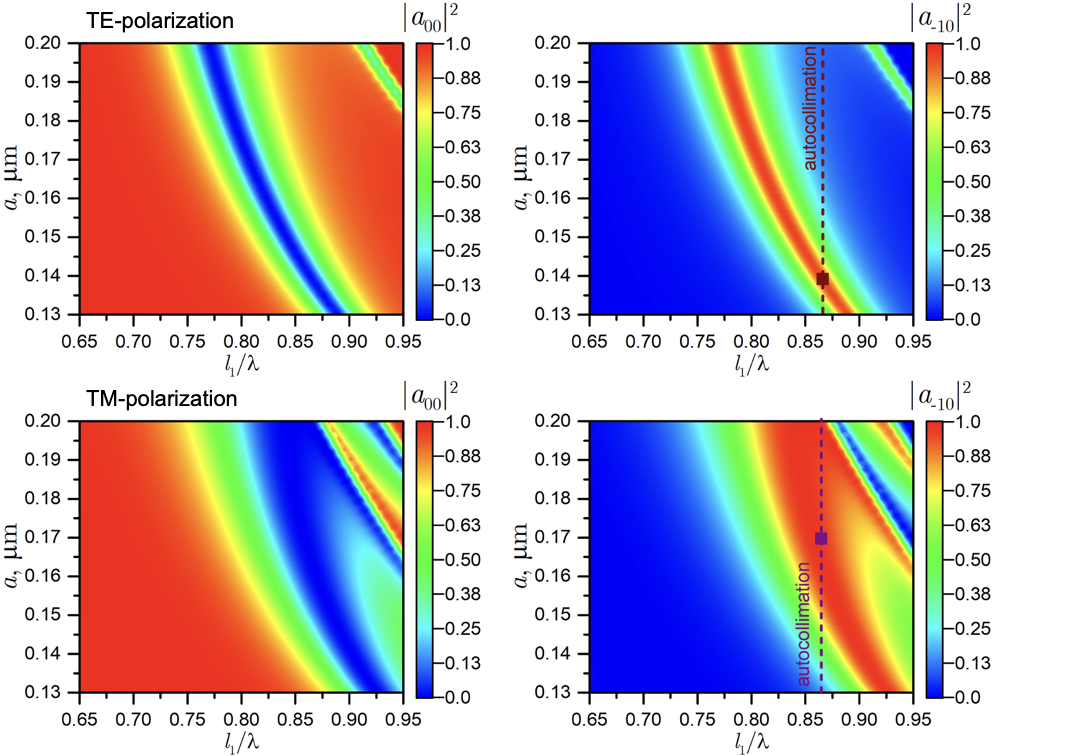}\\
  \caption{Dependencies of intensities $|a_{00}|^2$ (left pictures) and $|a_{-10}|^2$ (right pictures)
  corresponding to TE-polarization (upper row) and TM-polarization (bottom row) of incident wave are shown versus a disk radius $a$ and normalise frequency: $\alpha=\pi/2,$ $l_1=l_2=0.75$ $\mu$m,  $b=0.08$ $\mu$m, $c=0$, $\phi_i=0$, 
  and $\theta_{i}=35$ degrees. We marked the normalise frequency $l_1/\lambda=0.87$ corresponding to the Littrow scenario of diffraction by the dashed line and the disk radius provided maximum intensity of related non-specular reflection by the  black square point.}\label{a-f}
\end{figure}

\begin{figure}
\centering
  \includegraphics[height=4in]{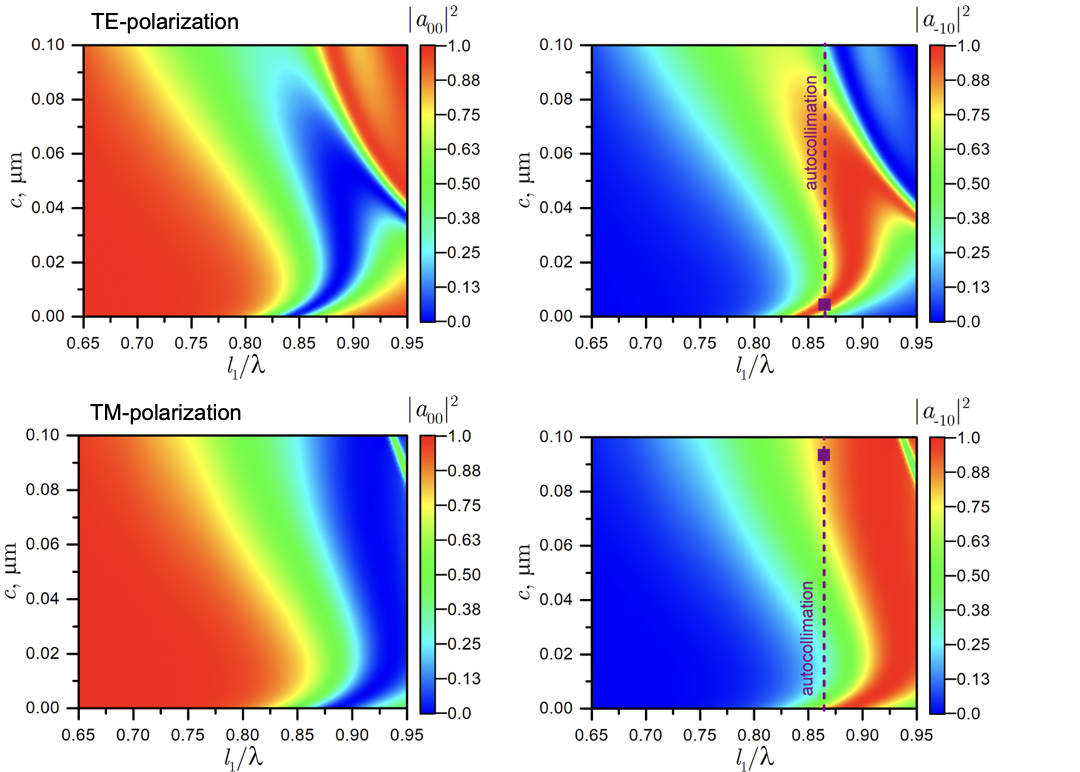}\\
  \caption{Dependencies of intensities $|a_{00}|^2$ (left pictures) and $|a_{-10}|^2$ (right pictures)
  corresponding to TE-polarization (upper row) and TM-polarization (bottom row) of incident wave are shown versus a thickness $c$ of substrate dielectric layer and normalise frequency:  $\alpha=\pi/2,$ $l_1=l_2=0.75$ $\mu$m,  $a=0.15$ $\mu$m, $b=0.08$  $\mu$m, $\phi_i=0$, and $\theta_{i}=35$ degrees. We marked the normalise frequency $l_1/\lambda=0.87$ corresponding to the Littrow scenario of diffraction by the dashed line and the thickness of substrate dielectric layer provided maximum intensity of related non-specular reflection by the black square point.}\label{c-f}
\end{figure}

In order to determine the range of metasurface parameters that provide full non-specular reflection of the incident wave, colour maps of the dependence of the reflectance $|a_{00}|^2$ and diffraction efficiency $|a_{-10}|^2$ on the disk radius, the thickness of dielectric layer of substrate and normalised frequency were simulated and shown in Fig. \ref{a-f} and Fig. \ref{c-f}. The amplitudes of partial waves of reflected field (\ref{E1}) are defined by equation  $|\bi{d}_{mn}|^2=|a_{mn}|^2 \gamma_{00}/k.$

The metasurface manifests resonant frequency dependencies of specular and non-specular reflection intensities for both TE- and TM-polarization of incident wave. The enlargement of disk diameter leads to the decrease of resonant frequency. The quality factor of resonance is essentially higher for the case of excitation by TE-polarize wave in comparison with the case of  TM-polarization. In any case of polarization, the metasurface provides an opportunity for full non-specular reflection both TE- and TM-polarize waves.

We also reach the opportunity of autocollimation diffraction regime with full intensity transformation into reversal wave. However, there is the need the optimisation over the whole set of problem parameters to derive the conditions for overlapping of autocollimation regimes of two polarizations. It was not the subject of this research.

In the presence of a dielectric layer between the disk array and the metal screen, the physical picture of reflection is changed due to changes in the conditions of resonance. As the thickness of non-resonant dielectric layer increases, the effect of the metal screen  decreases, that leads to increasing the frequency of $\mbox{HEM}_{11\delta}$-oscillation, and the oscillation frequency of  $\mbox{TE}_{01\delta}$  decreases (see Fig. \ref{c-f}). This feature gives us the opportunity to choose the thickness $c$ of dielectric layer to control the relative position of the resonances in the frequency scale, in particular, to achieve their overlapping. It should be noted that in the case of overlapping at least two resonances lead to the appearance of a wide frequency zone with effective non-specular reflection.

\begin{figure}
\centering
  \includegraphics[height=3in]{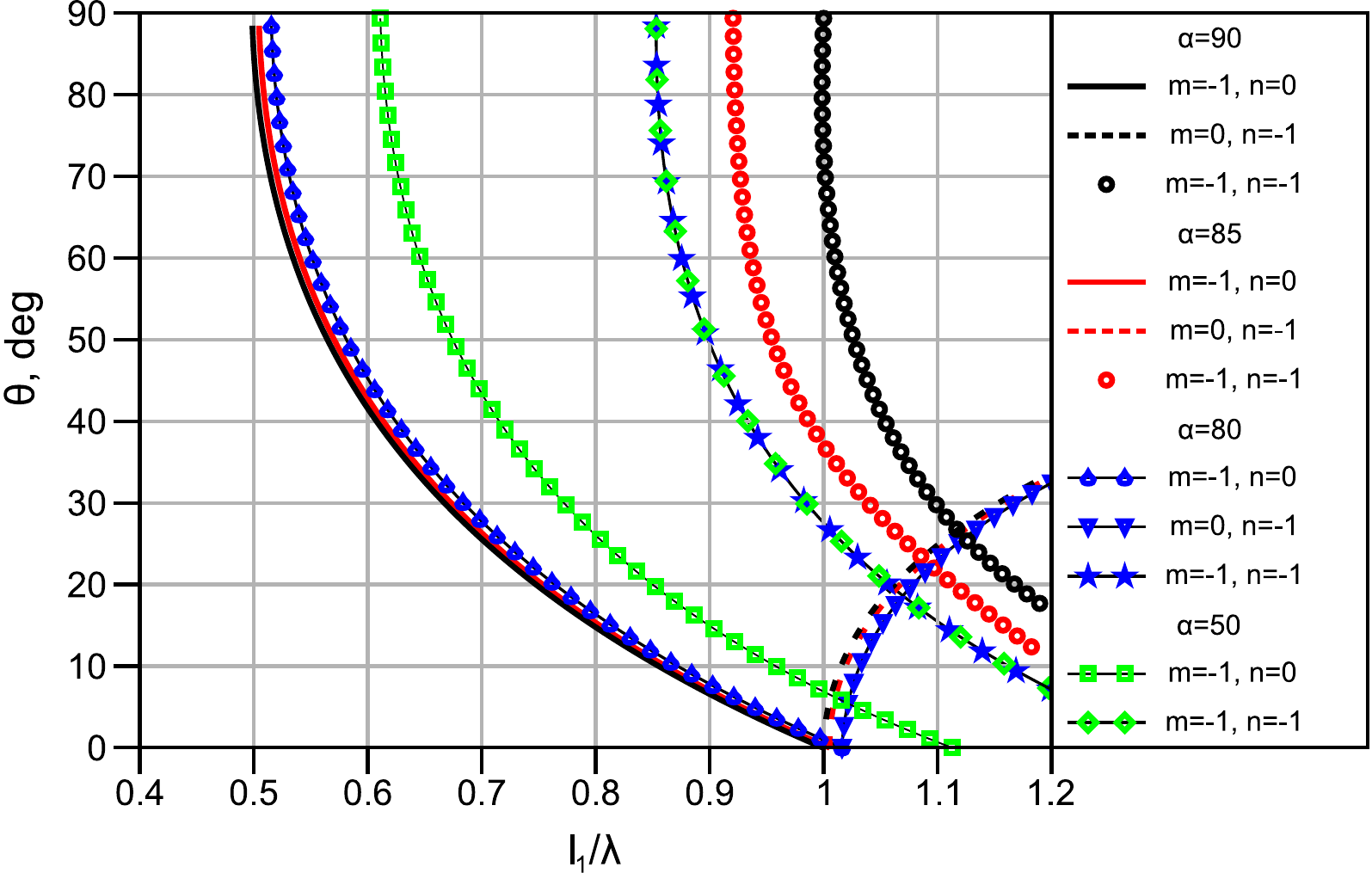}
  \caption{A diagram defined zones of parameters corresponded to existence of two non-evanescent partial waves (a specular 00-wave and an alone non-specular wave) in the reflected field of a periodic structure with a rhombic unit cell. The areas are bounded by two lines of different type and identical colour.} \label{InclZones}
\end{figure}

The non-orthogonality of the directions of array periodicity ($Os_1$ and $Os_2$ axes)  does not lead to significant changes in the resonant properties of the metasurface, as the latter ones are associated with the natural oscillations of the field inside the disks. Due to the disk symmetry, the variation of the angle between the periodicity directions can only change a disk coupling, but does not change the field structure of natural oscillations. On the other hand, if the angles of a rhombic unit cell ($l_1=l_2$) are smaller then $\alpha=60$ degrees, the first non-specular non-evanescent diffraction order becomes a ray (-1, -1) instead of (-1,0). This diffraction order propagates at an angle to the plane of incidence. In Fig. \ref{InclZones}, we show diagrams that can be used to track the location of an area in which there are only two diffraction rays in the reflected field versus changing the angle $\alpha$.

\begin{figure}
\centering
  \includegraphics[height=4.5in]{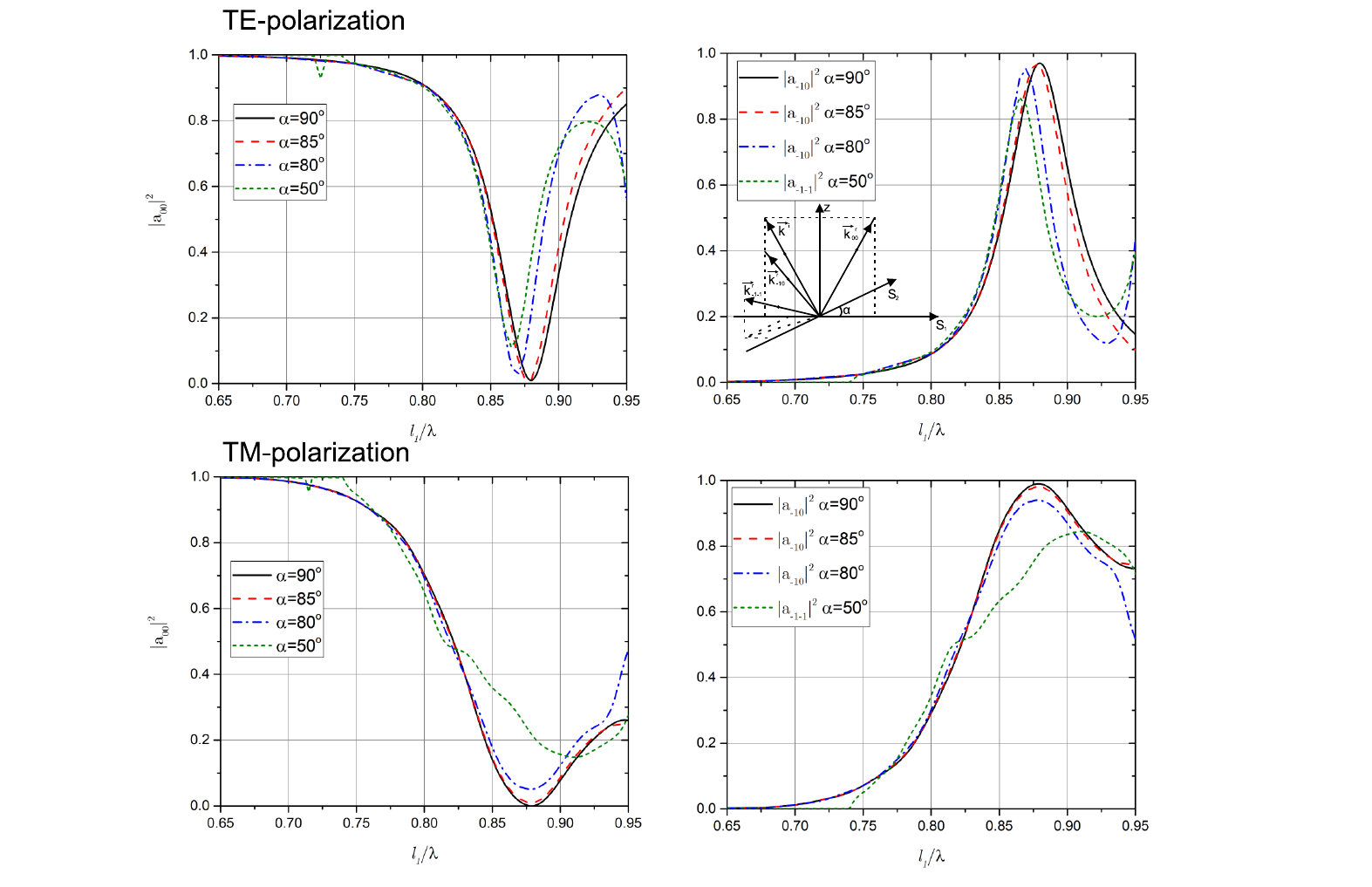}
  \caption{Frequency dependencies of reflection intensities of main diffraction order $|a_{00}|^2$ (left graphs) and  first partial waves ($|a_{-10}|^2$ and $|a_{-1-1}|^2$, see graphs in right column) corresponding to TE- (upper row, $a=0.15$ $\mu$m) and TM-polarization (bottom row, $a=0.175$ $\mu$m) of incident wave for the metasurface with a rhombic unit cell:   
  $l_1=l_2=0.75$ $\mu$m,  $b=0.08$ $\mu$m, $c=0.01$ $\mu$m, and $\phi_i=0,$ $\theta_i=35$ degrees. Upper figure of right column includes a scheme that demonstrate direction of propagation $(-1,0)$ and $(-1,-1)$ diffraction order in reflected field.}\label{rhomb}
\end{figure}

For comparison we present here frequency dependencies of coefficients of specular and non-specular reflection by metasurface with a rhombic-shaped unit cell of the array (see Fig. \ref{rhomb}). Resonances of full non-specular reflection are observed for wave diffraction by these kind of metasurface. It also should be noted that this kind of resonance may be manifested by diffraction order propagated outside of the plane of incidence.  

\section*{Conclusions}

In  this  paper  we  demonstrate  a  numerical  analysis  of  a  non-specular  reflection of  wave  radiation  by  the  silicon-on-metal  metasurface  involving  the  excitation  of  diffraction orders besides the fundamental one.  To provide a full non-specular reflection for any direction and any polarization state of incident wave, we propose for the first time a metamaterial mirror composed of silicon disks array with excitation of the Mie-resonances forming a specific symmetry of field distribution that can enhance or suppress specific diffraction orders resulting in anomalous reflection. This kind of mirror can be a key component in planar optoelectronic devices.

\ack
Authors are grateful to the National Research Fund of Ukraine for support of this work by the project grant 2020.02/0218.
% references section

\section*{References}
% NOTE: BibTeX documentation can be easily obtained at:
% http://www.ctan.org/tex-archive/biblio/bibtex/contrib/doc/

%\begin{thebibliography}{99}

\bibliographystyle{unsrt}

\bibliography{Step_1_NRFU_1}

\begin{thebibliography}{10}

\bibitem{ancient_mirrors-2006}
Jay~M. Enoch.
\newblock History of mirrors dating back 8000 years.
\newblock {\em Optometry and Vision Science}, 83:775--781, October 2006.

\bibitem{metasurface-2016}
Stanislav~B. Glybovski, Sergei~A. Tretyakov, Pavel~A. Belov, Yuri~S. Kivshar,
  and Constantin~R. Simovski.
\newblock Metasurfaces: From microwaves to visible.
\newblock {\em Physics Reports}, 634:1--72, 2016.
\newblock Metasurfaces: From microwaves to visible.

\bibitem{meta_mirrors-2014}
Majid Esfandyarpour, Erik~C. Garnett, Yi~Cui, Michael~D. McGehee, and Mark~L.
  Brongersma.
\newblock Metamaterial mirrors in optoelectronic devices.
\newblock {\em Nature Nanotechnology}, 9(7):542--547, 2014.

\bibitem{meng-2017}
Jingshi Meng, Trevon Badloe, Jungho Mun, and Junsuk Rho.
\newblock Metasurfaces-based absorption and reflection control: Perfect
  absorbers and reflectors.
\newblock {\em Journal of Nanomaterials}, 2017:2361042, 2017.

\bibitem{yablonovitch-2015}
Michael~S Eggleston, Kevin Messer, Liming Zhang, Eli Yablonovitch, and Ming~C
  Wu.
\newblock Optical antenna enhanced spontaneous emission.
\newblock {\em Proceedings of the National Academy of Sciences of the United
  States of America}, 112(6):1704--1709, 02 2015.

\bibitem{4666749}
D.~C. {Li}, F.~{Boone}, M.~{Bozzi}, L.~{Perregrini}, and K.~{Wu}.
\newblock Concept of virtual electric/magnetic walls and its realization with
  artificial magnetic conductor technique.
\newblock {\em IEEE Microwave and Wireless Components Letters},
  18(11):743--745, 2008.

\bibitem{all-dielectric-2016}
Saman Jahani and Zubin Jacob.
\newblock All-dielectric metamaterials.
\newblock {\em Nature Nanotechnology}, 11:23--36, 1 2016.

\bibitem{absorber_for_sensor_2015}
Ben-Xin Wang, Xiang Zhai, Gui-Zhen Wang, Wei-Qing Huang, and Ling-Ling Wang.
\newblock A novel dual-band terahertz metamaterial absorber for a sensor
  application.
\newblock {\em Journal of Applied Physics}, 117(1):014504, 2015.

\bibitem{multispectral_sensing_2015}
Riad Yahiaoui, Siyu Tan, Longqing Cong, Ranjan Singh, Fengping Yan, and Weili
  Zhang.
\newblock Multispectral terahertz sensing with highly flexible ultrathin
  metamaterial absorber.
\newblock {\em Journal of Applied Physics}, 118(8):083103, 2015.

\bibitem{sydorchuk-2017}
N~Sydorchuk and S~Prosvirnin.
\newblock Analysis of terahertz wave reflection by an array of double
  dielectric elements placed on a reflective substrate.
\newblock In {\em XXIInd Intern. Seminar/Workshop on Direct and Inverse
  Problems of Electromagnetic and Acoustic Wave Theory (DIPED)}, pages 58--63,
  Dnipro, Ukraine, September 2017.

\bibitem{sensors-2017}
Y.~Lee, S.~J. Kim, H.~Park, and B.~Lee.
\newblock Metamaterials and metasurfaces for sensor applications.
\newblock {\em Sensors}, 17:1726, Jul 2017.

\bibitem{Collin_2014}
St{\'{e}}phane Collin.
\newblock Nanostructure arrays in free-space: optical properties and
  applications.
\newblock {\em Reports on Progress in Physics}, 77(12):126402, nov 2014.

\bibitem{Zhu:15}
Li~Zhu, Jonas Kapraun, James Ferrara, and Connie~J. Chang-Hasnain.
\newblock Flexible photonic metastructures for tunable coloration.
\newblock {\em Optica}, 2(3):255--258, Mar 2015.

\bibitem{khardikov-2010-tol}
V.~V. Khardikov, E.~O. Iarko, and S.~L. Prosvirnin.
\newblock Trapping of light by metal arrays.
\newblock {\em J. Opt.}, 12:045102(11), March 2010.

\bibitem{khardikov-2012-jop}
V.~V. Khardikov, E.~O. Iarko, and S.~L. Prosvirnin.
\newblock A giant red shift and enhancement of the light confinement in a
  planar array of dielectric bars.
\newblock {\em J. Opt.}, 14:035103, 2012.

\bibitem{zheludev-2013-Near-infrared}
J.~Zhang, K.~F. MacDonald, and N.~I. Zheludev.
\newblock Near-infrared trapped mode magnetic resonance in an all-dielectric
  metamaterial.
\newblock {\em Optics Express}, 21(22):26721--26728, 2013.

\bibitem{prosvirnin-2009-jopa}
S.~L. Prosvirnin and N.~I. Zheludev.
\newblock Analysis of polarization transformations by a planar chiral array of
  complex-shaped particles.
\newblock {\em J. Opt. A: Pure Appl. Opt.}, 11:074002(10), 2009.

\bibitem{Gribovsky_2014}
Aleksander~V Gribovsky and Oleg~A Yeliseyev.
\newblock Nonspecular reflection of gaussian wave beams on a two-dimensional
  periodic array with shorted waveguides of rectangular cross-section.
\newblock {\em Journal of Optics}, 16(3):035701, Feb 2014.

\bibitem{Amitay-1972-tapaa}
Noach Amitay, Victor Galindo, and Cheng~Pang Wu.
\newblock {\em Theory and analysis of phased array antennas}.
\newblock Wiley - Interscience, a Division of John Wiley and Sons, Inc., New
  York-London-Sydney-Toronto, 1972.

\bibitem{collin-2001}
R.~E. Collin.
\newblock {\em Foundation for Microwave Engineering}.
\newblock Wiley, New York, second edition, 2001.

\bibitem{vandeGroep:13}
J.~van~de Groep and A.~Polman.
\newblock Designing dielectric resonators on substrates: Combining magnetic and
  electric resonances.
\newblock {\em Opt. Express}, 21(22):26285--26302, Nov 2013.

\bibitem{dielectric-antenna-2011}
M.~Ene-Dobre, Marian Banciu, Liviu Nedelcu, G.~Stoica, Cristina Busuioc, and
  Horia Alexandru.
\newblock Microwave antennas based on $\mathrm{Ba_{1-x}Pb_x Nd_2 Ti_5 O_{14}}$.
\newblock {\em Journal of Optoelectronics and Advanced Materials},
  13:1298--1304, 10 2011.

\bibitem{hard-1970}
T.~M. Hard.
\newblock Laser wavelength selection and output coupling by a grating.
\newblock {\em Appl. Opt.}, 9:1825--1830, Aug 1970.

\bibitem{Masalov-1980}
S.~A. Masalov and Iu.~K. Sirenko.
\newblock Excitation of reflecting lattices by a plane wave in the
  autocollimation mode.
\newblock {\em Radiophysics and Quantum Electronics}, 23(4):332--338, Oct 1980.

\bibitem{Lotem-1994}
Haim Lotem.
\newblock Littrow-mounted diffraction grating cavity.
\newblock {\em Applied Optics}, 33(6):930--934, Feb 1994.

\end{thebibliography}

%\end{thebibliography}

\end{document}